# Micron-sized domains in quasi single-component giant vesicles

Roland L. Knorr*, Jan Steinkühler and Rumiana Dimova*

Max Planck Institute of Colloids and Interfaces, Department of Theory & Bio-Systems, Science Park Golm, 14424 Potsdam, Germany

*Address correspondence to: knorr@mpikg.mpg.de and dimova@mpikg.mpg.de

**Abstract**

Giant unilamellar vesicles (GUVs), are a convenient tool to study membrane-bound processes using optical microscopy. An increasing number of studies highlights the potential of these model membranes when addressing questions in membrane biophysics and cell-biology. Among them, phase transitions and domain formation, dynamics and stability in raft-like mixtures are probably some of the most intensively investigated. In doing so, many research teams rely on standard protocols for GUV preparation and handling involving the use of sugar solutions. Here, we demonstrate that following such a standard approach can lead to abnormal formation of micron-sized domains in GUVs grown from only a single phospholipid. The membrane heterogeneity is visualized by means of a small fraction (0.1 mol%) of a fluorescent lipid dye. For dipalmitoylphosphatidylcholine GUVs, different types of membrane heterogeneities were detected. First, unexpected formation of micron-sized dye-depleted domains was observed upon cooling. These domains nucleated about 10 K above the lipid main phase transition temperature, $T_M$. In addition, upon further cooling of the GUVs down to the immediate vicinity of $T_M$, stripe-like dye-enriched structures around the domains are detected. The micron-sized domains in quasi single-component GUVs were observed also when using two other lipids. Whereas the stripe structures are related to the phase transition of the lipid, the dye-excluding domains seem to be caused by traces of impurities present in the glucose. Supplementing glucose solutions with nm-sized liposomes at millimolar lipid concentration suppresses the formation of the micron-sized domains, presumably by providing competitive binding of the impurities to the liposome membrane in excess. It is likely that such traces of impurities can significantly alter lipid phase diagrams and cause differences among reported ones.

## 1 Introduction

Increasing interdisciplinary attention to multicomponent model membranes is raised by the discovery of nano- and micron-scale phase separation occurring in cellular membranes and plasma membrane mimetics such as blebs (also called giant plasma membrane vesicles) [1-5]. Proteins partitioning between two coexisting fluid phases (liquid ordered and liquid disordered) is believed to play a role in cellular signaling and sorting processes. One important result of the search for evidence for cell membrane phase separation was the discovery of phase separation in model membranes made of ternary lipid mixtures, see e.g. Refs. [6-9]. This phenomenon can be directly observed in giant unilamellar vesicles (GUVs) [10-13] using fluorescence microscopy. Ternary lipid mixtures are well suited model systems to study lipid-driven phase separations on the nano- and micro scale. Such mixtures contain cholesterol, a phospholipid with a low phase transition temperature (lower than the experimental temperature of observation) and sphingomyelin or a phospholipid with high transition temperature (typically above 35°C). In the region between these two phase transition temperatures, phase separation in the membrane of giant vesicles can be observed directly under the microscope. Most physiologically relevant are the two fluid phases - liquid ordered ($l_o$) and liquid disordered ($l_d$), although solid (or gel) phases are also present in skin and lungs. Phase separation can be observed also in two-component membranes composed of a high and a low melting temperature phospholipids, where the process leads to the coexistence of solid and fluid domains. Membrane-bound molecules partition differently between the phases and thus, membrane fluorescent dyes are efficient tools to image different phases [14, 15]. Concentrations of membrane dyes in such experiments typically range between 0.1 and 3 mol%, the higher fractions generally needed for epi-fluorescence microscopy, and the lower, being sufficient for confocal microscopy imaging. Because of the low fraction employed, membrane dyes are in practice not accounted for when distinguishing the membrane components such that binary mixtures (of two lipids) are in reality quasi-binary when a lipid dye is included; similarly, labeled GUVs made of one lipid and a dye are quasi single-component. Ignoring the dye molecule might be justified when it is not affecting the examined membrane properties, even though it might influence the phase separation process as observed from the shape of microscopic domains [16].

Here, we explore single-component GUVs, doped with a small fraction of a fluorescent lipid dye. The vesicles are subjected to heating and cooling. To our surprise, we find that micron-sized domains can form well above the main phase transition temperature of the lipid. The domains persist and are observed also at lower temperatures. This result was confirmed using different phosphatidylcholines. A number of various reasons for the generation of the domains are investigated, followed by a discussion on the implications of these finding to studies employing giant vesicles for investigating phase separation and domain formation in membranes.

## 2 Materials and Methods

### 2.1 Vesicle preparation

For the vesicle preparation, we employed the following lipids dissolved in chloroform: dipalmitoylphosphatidylcholine (DPPC, Avanti Polar Lipids, Birmingham AL), distearoylphosphatidylcholine (DSPC, Avanti), stearoyloleoylphosphatidylcholine (SOPC, Avanti) or a mixture of egg sphingomyelin (eSM, Avanti) and cholesterol (Sigma-Aldrich, Munich). Occasionally, as specified in the text, we employed D-α-phosphatidylcholine, dipalmitoyl (D-DPPC, Sigma-Aldrich) and DL-α-phosphatidylcholine, dipalmitoyl (D/N-DPPC, Sigma-Aldrich). GUVs were prepared following the electroformation method [17] with modifications as described in Ref. [18]. Briefly, the vesicles were grown from a lipid film deposited on conductive glasses (coated with indium tin oxide). (Occasionally, electroformation was performed on platinum wires using a home-made chamber.) After spreading the lipid-chloroform solutions on the electrodes, the glasses were kept for 2 h under vacuum and subsequently assembled into a chamber with a 2 mm Teflon spacer. The chamber was then filled with sucrose solution or Milli-Q water. The chamber was held together by binder clips to avoid the use of sealants such as silicone grease which might affect the membrane properties [19]. The chamber electrodes, i.e. the conductive glasses, were connected to a function generator and alternating current of 1 V (root-mean squared) and 10 Hz frequency was applied for ~ 2 h at 60°C. After electroformation, the chambers were cooled down to room temperature (23°C) at the rates indicated in the text, the vesicles were harvested and stored at room temperature until use (within a day). Spontaneous swelling was performed from lipid films spread on bare glass cover slips in 200 mM sucrose over night at 60°C without prehydration. In both preparation protocols, electroformation and spontaneous swelling, the formation chambers were placed on a large aluminum block inside the oven at 60 °C which reduced drastic temperature changes during sample manipulation after vesicle preparation.

All GUVs were labeled with 0.1 mol% fluorescent dye dissolved in the chloroform lipid stock solution. The following dyes were used: 1,2-dioleoyl-sn-glycero-3-phosphoethanolamine-N-(lissamine rhodamine B sulfonyl) (ammonium salt) (Rh-DPPE, Avanti) or 1,1'-dioctadecyl-3,3,3',3'- tetramethylindotricarbocyanine Iodide (DiIC18, Thermofisher, Waltham, MA ). Occasionally, as specified in the text, we used 1,2-dipalmitoyl-sn-glycero-3-phosphoethanolamine-N-(7-nitro-2-1,3-benzoxadiazol-4-yl) (DPPE-NBD, Avanti), 1,2-dihexadecanoyl-sn-glycero-3-phosphoethanolamine, triethylammonium salt (TR-DHPE, Invitrogen) and perylene (Sigma-Aldrich).

Single-component GUVs grown in sucrose were diluted at 1:10 ratio with an isoosmolar glucose solution. Sucrose and glucose were obtained from Sigma-Aldrich (BioUltra > 99.5 % purity by HPLC) and for a few tests, from Fluka. GUVs prepared from eSM and cholesterol (Chol) were diluted at 2:1 ratio in an isoosmolar glucose solution. Slightly hypertonic and hypotonic glucose solutions (180 mM and 220 mM instead of 200 mM) were used for controls.

Small vesicles from SOPC or DPPC at a final lipid concentration of 10 mM were prepared by six freeze-thaw cycles with liquid nitrogen in 200 mM glucose.

### 2.2. Temperature control chamber and vesicle observation

Various simple observation chambers were tested, home build from cover slips, silicone grease and/or press-to-seal spacers (Sigma). All gave similar results. The temperature decay inside the observation chamber was monitored with a fiber-optic temperature probe attached to a signal conditioner (FOT-M and FTI-10, FISO Technology, Canada) with an accuracy of +/- 0.01 K. For precise control over the temperature, we built a chamber as shown in Fig. 1, see also Ref. [20]. The chamber was formed by fixing a microscope slide to an aluminum block with epoxy glue. The aluminum block had windows for illuminating the sample for transmission light microscopy. The temperature of the aluminum block was controlled by circulating heating/cooling water connected to a thermostat (Neslab RTE, Portsmouth, NH). The temperature variation in the chamber as a function of the distance from chamber bottom was less than 0.5 K [20].

For confocal and phase contrast imaging, the observation chamber was mounted on a Leica SP5 system (Mannheim, Germany) equipped with a 40x HCX Plan APO objective (NA 0.75). The vesicles labeled with Rh-DPPE or DiIC18 were imaged using a diode-pumped solid-state laser at 561 nm for excitation and the emission signal was collected in the wavelength range from 575 nm to 650 nm. Alternatively, GUVs were observed under epi-fluorescence mode on an inverted microscope (Axio Observer D, Zeiss, Jena) equipped with a 40x, 0.6 NA objective using the appropriate filter sets.

## 3 Results

**Vesicle stability is reduced upon fast cooling after preparation**

The main phase transition temperature, $T_M$, of dipalmytoylphosphatidylcholine (DPPC) is 41°C. Below this temperature, the membrane can be in the crystalline ripple ($P_{\beta}'$) or planar ($L_{\beta}'$) phase. In the following we do not distinguish the various polymorphs and refer to all non-fluid phases as gel. We grew the GUVs at 60 °C to ensure lipid fluidity thus enabling GUV formation from the planar bilayer stacks. At this temperature, the preferred methods for GUV preparation are electroformation and spontaneous swelling (for literature on GUV preparation methods, see e.g. Refs. [12, 21, 22]). Approaches based on vesicle swelling on polymer films [23, 24] were not employed because they result in polymer residues encapsulated in the vesicles and maybe even in the membrane [25, 26], which is more pronounced at high temperature. Methods based on transfer of lipids from an oil phase to an aqueous one [27, 28] were also avoided because of the inherent risk of remaining oil residues in the bilayer, which may influence the membrane phase behavior.

In conventional preparation approaches, vesicles are typically grown in sucrose solution and subsequently diluted in glucose. In this way, the GUVs settle to the bottom of the observation chamber because of the density difference of the sugar solutions. This greatly facilitates observation. Here, if not otherwise stated, the GUVs were prepared in 200 mM sucrose solution. After slowly cooling the samples (0.15 K/min, comparable to and even slower than cooling rates typically used in differential scanning calorimetry) to room temperature, the GUVs were diluted at 1:10 ratio in isoosmolar glucose solution. With this sugar contrast, GUVs appear as dark objects with bright halo when observed under phase contrast microscopy (see e.g. right vesicle in Fig. 2C) because of the differences in the refractive index of the sucrose and glucose solutions. The enhanced phase contrast across the membrane allows distinguishing vesicles which have leaked (with lost contrast) from GUVs with intact membrane (i.e. without pores); see Fig. S1A in the Supporting Information (SI). The phase contrast of the majority of the vesicles was preserved, indicating that in these vesicles, membrane pores were absent.

We examined the effect of different cooling rates on the vesicle size and stability. Increasing the cooling rate from 0.15 K/min to 4 K/min resulted in overall reduced vesicle size and the membrane of a significant number of the vesicles appeared to have been compromised as evidenced by loss of contrast indicating pore formation, Fig. S1B. By fast cooling, most large GUVs lost contrast. Further, we worked with vesicles in slowly cooled samples only. Confocal microscopy observation of such vesicles showed that the membrane surface appears inhomogeneous and grainy, see Section S1 in the SI. Such inhomogeneous fluorescence may be caused by dye partitioning and local membrane corrugations. Note that this inhomogeneity is difficult to detect with epifluorescence imaging because of the poorer resolution compared to confocal microscopy.

### Micron-sized domains in DPPC-GUVs vesicles appear after reheating

We then examined the response to reheating of GUVs electroformed in sucrose at 60 °C, slowly cooled to room temperature and diluted 1:10 in glucose. A closed chamber made of two cover slips was filled with 0.1 ml of the vesicle suspension, heated in an oven to 60 °C, i.e. well above the lipid melting temperature, and cooled to room temperature. To our surprise, we observed the formation of many "black" (dye-free), areas in the membrane of the GUVs, see Fig 2. Their typical diameters were in the range of 2-3 μm. In principle, such black regions could represent areas without a membrane, i.e. pores in the bilayer [18, 29]. Such large pores would equilibrate the glucose/sucrose asymmetry across the bilayer within minutes and result in a loss of phase contrast asymmetry. However, the black areas were observed also in GUVs with preserved phase contrast, Fig. 2C. This suggests that these areas were intact membranes excluding the fluorescent dye, i.e. membrane domains; see also Fig. S4 in the SI for additional images.

### Domains in DPPC-GUVs form well above the main phase transition of the lipid

In an attempt to observe the appearance of the domains in the GUVs, we heated the vesicles to 60°C, placed the sample immediately on the microscope stage and observed them while slowly cooling. To achieve the slow cooling rate and to be able to record the temperature decay in the sample, we used a chamber with a larger volume of about 1 ml.

In the first minute after removing the chamber from the oven, the fluorescence signal was homogeneously distributed within the membrane of the GUVs (see e.g. Fig. S2B). Moreover, all vesicles fluctuated which indicated that the membrane was in the fluid phase as expected. Then, surprisingly, well above the main phase transition temperature, we observed nucleation of circular domains in the GUVs which excluded the fluorescent dye, Fig. 3A. The domains nucleated over the whole vesicle surface during cooling and grew over time. They were mobile and in some cases coalesced with each other, which suggested fluidity of both phases (at least around 50 °C, see section on domain coalescence further below). Temperature measurements in the observation chamber indicate that the black domains nucleated between 50 °C and 55 °C, Fig. 3C, well above the main phase transition temperature of DPPC (more details on the accuracy of temperature measurements are found in [20]). Occasionally, some domains attained unusual heart-like shapes (Fig. 3A) or exhibited kinks in the boundary, both reminiscent of patterns of gel domains observed in lipid monolayers for example [30, 31]. Further during cooling, the contact area of the GUVs with the bottom of the observation chamber decreased and became more circular (see the evolution of the red region between 9 s and 33 s in Fig. 3A), suggesting that the excess area of the vesicle has decreased and the GUVs became more spherical and tenser. This is presumably associated with a reduction of the area per molecule during cooling. This observation is in agreement with a study on changes of the adhesion area of DPPC GUVs near the main phase transition [32]. Interestingly, in one of the vesicle images published in Ref. [32], similar domains as those we report here are visible, but they were not discussed. The reduction of the contact area of the vesicle with the substrate also speaks about no or weak adhesion of the vesicles, supported by observations of their vertical cross sections as in [33]. We thus can exclude that the domain formation resulted from adhesion as reported in other systems [34, 35].

### Stripe-like structures in DPPC GUVs form close to the main phase transition

About 4 minutes after initiating the cooling, and in addition to the black domains, stripe-like structures appeared in the membrane surrounding the domains, Fig. 3B. The stipes appeared enriched with fluorescent dye and encircled the black domains. The temperature was between 42°C and 43°C, i.e. in the immediate vicinity of the main phase transition temperature $T_M$, Fig. 3C. Within 20-30 seconds after the initial appearance of the stripes, the relative movements of the black domains (and stripes) was arrested suggesting gelation of the dye-rich phase, the vesicles often adopted a faceted and edgy shape and many GUVs ruptured spreading on the cover slip. The thermal expansion coefficient for the volume of water is small compared to the thermal expansion coefficient for the area of lipid bilayers (examples of temperature-induced shape transformations as a function of temperature but in fluid GUVs is provided in Ref. [36]), thus a decrease in temperate shrinks the vesicle area much more than its volume (which is equivalent to building tension in the membrane) and may lead to rupture.

The complex surface patterns of dye-depleted domains and stripe-like structures which formed on the DPPC GUVs were stable for many days at room temperature. The stripe-like structures were observed in all GUVs of a given population. Their appearance coincides with the main phase transition on the GUV membrane, which has been reported to be $T_M(GUVs) = 41.7 \pm 1.5°C$, i.e. slightly different from that of systems with higher curvature or (multi)lamellarity [37]. Note that $T_M(GUVs)$ in Ref. [37] was measured for similar sugar conditions (the vesicles were prepared in 100 mM sucrose and diluted in isotonic glucose). The appearance of the micronized black domains was unexpected which motivated us to study the conditions of their formation further.

### Domain coalescence and formation is reversible by changes in temperature

To probe the fluidity and stability of the black domains, we arrested the temperature shortly after observing their appearance (in epifluorescence), and then reheated the vesicles, Fig 4A. For this purpose, we used a temperature-control chamber, see Fig. 1 and Materials and Methods. Whereas at 52°C the majority of the GUVs showed homogeneous distribution of the fluorescent dye, after a temperature quench to 49°C the majority appeared phase separated and multiple small dye-excluding domains were detected, Fig 4A left to right. When the temperature was again increased to 52°C, the domains dissolved again. When left at 49°C, Fig. 4B, the boundary of the domains was observed to visibly fluctuate, indicting fluidity. The domains grew by coalescence, although at extremely slow rate suggesting high viscosity of the membrane in the domain (note that solid domains should follow constant growth kinetics, resulting in uniform size while we observe domains of varying diameter Fig S5). The continuous (fluorescent) phase, on the contrary appeared to be more fluid as observed by the diffusion of individual domains over time. Presumably, increase in the viscosity and solidifying of the domains with lowering temperature cannot be excluded and need further investigation.

We were able to measure the nucleation temperature of the circular domains more precisely. By temperature cycling and monitoring the formation and dissolution of the fluid domains on the same GUV repeatedly, the miscibility temperature was found to be between 51°C and 52°C, i.e. the nucleation temperature of the domains is about 10 K above $T_M$. Their appearance is reversible, which implies that they reconstitute a thermodynamic phase and that the domains are not resulting from photooxidation.

**Formation of micron-sized domains is independent on lipids, dye and methods**

To exclude artifacts associated with the used chemicals or methods, we tested whether the formation of fluid domains upon cooling is related to the specific batch of DPPC, fluorescent dye and solutions. The experiments are summarized in Table S1. We used different batches of DPPC from Avanti, DPPC of different chirality and produced by a different manufacturer (D-DPPC, D/L-DPPC from Sigma). Moreover we tested other membrane dyes (NBD-DPPE, TR-DHPE, and perylene) as well as sucrose and glucose from other producers (Fluka and Sigma). Hyper- or hypoosmotic conditions were also explored. In all of the tested cases, we could still observe formation of black domains on the surface of the GUVs. We also found that neither the method of GUV formation (electroformation on glasses with indium tin oxide coating or platinum wires; spontaneous swelling) nor the type of observation chamber influenced the result.

We also tested two other phosphatidylcholines for domain formation: DSPC and SOPC. Similarly to DPPC GUVs, we prepared vesicles in 200 mM sucrose by electroformation about 20 K above the respective main phase transition temperatures of the lipids $T_M$(DSPC) = 55°C and $T_M$(SOPC) = 6°C. We then cooled them to room temperature, diluted them in isotonic glucose solution, and subjected them to one temperature cycle across their respective $T_M$ (the SOPC GUVs were cooled on ice to 0 °C) before observing them at room temperature. GUVs from both lipids exhibited dye-excluding micron-sized domains, similar to those in DPPC GUVs, Fig. 5. However whereas the shape of the domains in the DPPC vesicles in the gel phase appeared circular, irregular or hexagonal in some cases (Figs. 5B and S4), the domains in the DSPC GUVs were diamond-shaped (Fig. 5A). At room temperature SOPC is fluid, and the domains were of irregular shape and smooth boundaries, Fig. 6C. Stripe-like structures with higher fluorescence intensity as in the gel DPPC vesicles were not observed in GUVs grown from DSPC or SOPC. In summary, we found that the formation of the dye-excluding domains is very robust and persistent under a broad range of standard experimental conditions, and various lipids.

**Micron-sized domains do not form in the absence of glucose and can be suppressed by increasing the lipid concentration**

Interestingly, we noticed that reheating DPPC GUVs grown in pure water or sucrose (without any dilution in glucose) does not result in the formation of black domains as those shown in Figs. 3-5, see Table S1. Instead, the vesicles appeared to have inhomogeneous surface structure as in Fig. S3 without dye-excluding micron-sized domains.

Imaging GUVs with glucose-sucrose solution asymmetry is a standard approach conventionally employed in many applications. The glucose we used is ultrapure (> 99.5 % purity by HPLC). Presumably, impurities (maximum of 0.5 %), corresponding to at most 1 mM equivalent concentration are introduced in the GUV suspension upon 1:10 dilution in 0.2 M glucose. The lipid concentration in our samples is about three orders of magnitude lower, i.e., in the micromollar range. Thus, even a low percentage of impurities binding to the membrane could be sufficient vesicles in practice to convert the GUVs into multicomponent and thus, to influence the phase behavior significantly. We speculated that the binding of putative impurities to GUVs may be reduced by providing an alternative membrane surface for binding. We thus supplemented the 0.2 M glucose solution (used to dilute the GUVs) with millimolar concentrations of small, unlabeled vesicles (SUVs) made of SOPC or DPPC. In this way, we provided an excess of membrane with chemically identical surface for binding the impurities. As expected, this strongly suppressed the formation of black domains, Fig. 6. Similarly, domain formation was absent in GUV samples grown in 15 mM sucrose and diluted in isotonic glucose which was supplemented with 1 mM SUVs, Fig. S7.

**Phase separation in binary lipid mixtures can also be caused by glucose**

Finally, we extended our studies to binary mixtures of egg sphingomyelin (eSM) and cholesterol (Chol). At Chol fractions between ~10 mol% and 20–30 mol%, studies on sphinogomyelin/Chol membranes show discrepancies about the phase state of the bilayer, see e.g. the phase diagram in Ref. [38], some studies suggesting no phase coexistence [39] and others reporting domain coexistence [40-42] in this range. Here, GUVs made of eSM/Chol 7/3 grown in water at 60°C and cooled down to room temperature appeared homogenous, Fig. S8A. In contrast when GUVs of the same composition were grown in sucrose and subsequently diluted in glucose solution in 2:1 ratio, micron-sized finger-like domains were observed, Fig. S8B. The domains retained their shape suggesting that they are in the solid (gel) phase, but (when not percolating the whole GUV) could displace relative to each other suggesting that the fluorescent phase was fluid. The appearance of these domains in the presence of glucose suggests that this sugar can affect the phase transitions of both, single and multicomponent membranes.

**4. Discussion and conclusion**

The occurrence of dye-depleted domains in two-component membranes is well-known in the literature; the domain shape is found to depend on membrane composition (particular lipid mixture) and membrane tension [43-45]. However, according to the Gibbs phase rule, single-component systems can exhibit only one homogenous phase at equilibrium. Thus, our results on phase separation in GUVs grown from a single lipid must be caused by an additional components present in the system, similar to the effect of buffers which have been observed to affect the phase state of the membrane [46]. Surely, the dye present in our vesicles is an additional membrane component. However, we observe that the micron-sized domains do not result from its presence and, in this sense, the membrane can be considered as single-component one. The observed black domains behave as a true equilibrium phase, i.e. exhibit a defined nucleation temperature and are generated independently of the cooling rate. Even more strikingly, the domains are observed also on membranes made of lipids with various acyl chain lengths (Fig. 5). We could exclude artifacts associated with the GUV preparation method (electroformation or spontaneous swelling both yield black domains) and chambers used for GUV observation. We find that the appearance of domains is linked to the introduction of glucose into the outside solution followed by a temperature cycle through the main phase transition (Fig. 3). In principle, the direct interaction between glucose and lipids as observed in molecular dynamics simulations (although at very high sugar concentration) [47] could in principle drive phase separation but, to our knowledge, no reports are available about this for the concentrations studied here. Sugars have been observed to affect the membrane bending rigidity of giant vesicles (see e.g. Ref. [48] and references therein) which contradicts data collected on bilayer stacks as summarized in [49, 50]. Note however, that the lipid-to-sugar concentrations in experiments with GUVs and with lipid stacks are orders away from each other, so are the lipid-to-impurities ratios, which might be a source for the observed discrepancy. Thus, when additional lipid material is supplied in the form of SUVs, as in our experiments here, the appearance of domains is largely suppressed (Figs. 6B,C and S7). Note that even in this case, glucose is still in excess (~200 mM glucose vs 1-10 mM lipids). We conclude that the observed fluid domains are stabilized by tentative impurities in commercially obtained glucose. We cannot speculate about the type of impurities that could affect the membrane phase state, but traces of calcium ions, could potentially play a role: calcium chloride has been show to shift the phase transition of membranes made of lipid mixtures containing charged lipids to higher temperatures and increase their tension [51], although the binding affinity to phosphatidylcholines is lower and this effect might be negligible. At this stage, our results indicate that studies employing sugar solutions at very high concentrations as e.g. in Refs. [52, 53] but also those we employ in this work (15-200 mM) should be performed with increased awareness of traces of impurities in commercially available sugars that might affect the phase behavior of the membrane, in addition to the effect of sugars themselves [54]. Further work will be required to quantify differences between sugars different producers and identify the membrane-binding component, which alters the phase behavior.

The low micromolar concentration of lipids in GUV preparations makes them particularly sensitive to micro- to millimolar concentrations of impurities. Experiments are often performed in sugar solutions or physiological buffers of concentrations in the 100-300 mM range. Typical specifications of chemicals from commercial producers allow between 0.5-5 % impurities. If some of these impurities have high affinity to the membrane, they may naturally influence its mechanical and thermodynamic properties. The presence of impurities in sucrose has been already documented and systematically found across manufactures [55]. Glucose is a standard solute in GUV experiments and tentative impurities could explain discrepancies between experiments conducted on GUVs and other lipid membrane systems (e.g. bilayer stacks) [50] or maybe even some discrepancies in GUV experiments performed in different labs. For example, phase diagrams of ternary mixtures reveal differences in the phase boundaries [38, 56]. Previously, we have shown that salt asymmetry can lead to changes in the phase diagram [20, 57]. Here, we demonstrate that addition of sugars to initially one-phase 7:3 eSM:Chol GUVs can also result in domain formation.

In summary we reported formation of micron-sized domains in single-component GUVs. Domain formation was observed by confocal and epifluorescence microscopy. We showed that the presence of the domains is very robust and does not depend on sample preparation or handling but on the presence of glucose – a standard chemical used in countless publications reporting work with GUVs.

**Acknowledgements:** This work is part of the MaxSynBio consortium which is jointly funded by the Federal Ministry of Education and Research of Germany and the Max Planck Society. We thank Reinhard Lipowsky (MPI of Colloids and Interfaces) for stimulating discussions, institutional and financial support.

**References**

[1] S.P. Rayermann, G.E. Rayermann, C.E. Cornell, A.J. Merz, S.L. Keller, Hallmarks of Reversible Separation of Living, Unperturbed Cell Membranes into Two Liquid Phases, Biophys J, 113 (2017) 2425-2432.
[2] T. Baumgart, A.T. Hammond, P. Sengupta, S.T. Hess, D.A. Holowka, B.A. Baird, W.W. Webb, Large-scale fluid/fluid phase separation of proteins and lipids in giant plasma membrane vesicles, Proceedings of the National Academy of Sciences, 104 (2007) 3165-3170.

[3] M. Carquin, L. D'Auria, H. Pollet, E.R. Bongarzone, D. Tyteca, Recent progress on lipid lateral heterogeneity in plasma membranes: From rafts to submicrometric domains, Prog. Lipid Res., 62 (2016) 1-24.
[4] C. Eggeling, C. Ringemann, R. Medda, G. Schwarzmann, K. Sandhoff, S. Polyakova, V.N. Belov, B. Hein, C. von Middendorff, A. Schonle, S.W. Hell, Direct observation of the nanoscale dynamics of membrane lipids in a living cell, Nature, 457 (2009) 1159-U1121.
[5] E. Klotzsch, G.J. Schutz, A critical survey of methods to detect plasma membrane rafts, Philos T R Soc B, 368 (2013).
[6] C. Dietrich, L.A. Bagatolli, Z.N. Volovyk, N.L. Thompson, M. Levi, K. Jacobson, E. Gratton, Lipid rafts reconstituted in model membranes, Biophys. J., 80 (2001) 1417-1428.
[7] S.L. Veatch, S.L. Keller, Separation of Liquid Phases in Giant Vesicles of Ternary Mixtures of Phospholipids and Cholesterol, Biophys J, 85 (2003) 3074-3083.
[8] J. Zhao, J. Wu, F.A. Heberle, T.T. Mills, P. Klawitter, G. Huang, G. Costanza, G.W. Feigenson, Phase studies of model biomembranes: Complex behavior of DSPC/DOPC/Cholesterol, Biochimica et Biophysica Acta (BBA) - Biomembranes, 1768 (2007) 2764-2776.
[9] C.C. Vequi-Suplicy, K.A. Riske, R.L. Knorr, R. Dimova, Vesicles with charged domains, Biochimica et Biophysica Acta (BBA) - Biomembranes, 1798 (2010) 1338-1347.
[10] R. Dimova, S. Aranda, N. Bezlyepkina, V. Nikolov, K.A. Riske, R. Lipowsky, A practical guide to giant vesicles. Probing the membrane nanoregime via optical microscopy, J. Phys.: Condens. Matter, 18 (2006) S1151-S1176.
[11] R. Dimova, Giant Vesicles: A Biomimetic Tool for Membrane Characterization, in: A. Iglič (Ed.) Advances in Planar Lipid Bilayers and Liposomes, Academic Press, Place Published, 2012, pp. 1-50.
[12] P. Walde, K. Cosentino, H. Engel, P. Stano, Giant Vesicles: Preparations and Applications, ChemBioChem, 11 (2010) 848-865.
[13] S.F. Fenz, K. Sengupta, Giant vesicles as cell models, Integr Biol-Uk, 4 (2012) 982-995.
[14] T. Baumgart, G. Hunt, E.R. Farkas, W.W. Webb, G.W. Feigenson, Fluorescence probe partitioning between L-o/L-d phases in lipid membranes, Biochim. Biophys. Acta-Biomembr., 1768 (2007) 2182-2194.
[15] Andrey S. Klymchenko, R. Kreder, Fluorescent Probes for Lipid Rafts: From Model Membranes to Living Cells, Chem. Biol., 21 (2014) 97-113.
[16] J. Juhasz, J.H. Davis, F.J. Sharom, Fluorescent probe partitioning in GUVs of binary phospholipid mixtures: Implications for interpreting phase behavior, Biochimica et Biophysica Acta (BBA) - Biomembranes, 1818 (2012) 19-26.
[17] M.I. Angelova, D.S. Dimitrov, Liposome electroformation, Faraday Discussions of the Chemical Society, 81 (1986) 303-311.
[18] R.L. Knorr, M. Staykova, R.S. Gracia, R. Dimova, Wrinkling and electroporation of giant vesicles in the gel phase, Soft Matter, 6 (2010) 1990-1996.
[19] G. Niggemann, M. Kummrow, W. Helfrich, The Bending Rigidity of Phosphatidylcholine Bilayers - Dependences on Experimental-Method, Sample Cell Sealing and Temperature, J. Phys. II, 5 (1995) 413-425.
[20] B. Kubsch, T. Robinson, J. Steinkuhler, R. Dimova, Phase Behavior of Charged Vesicles Under Symmetric and Asymmetric Solution Conditions Monitored with Fluorescence Microscopy, J. Vis. Exp, (2017) e56034.
[21] R. Dimova, S. Aranda, N. Bezlyepkina, V. Nikolov, K.A. Riske, R. Lipowsky, A practical guide to giant vesicles. Probing the membrane nanoregime via optical microscopy, J Phys Condens Matter, 18 (2006) S1151-1176.
[22] A.P. Liu, D.A. Fletcher, Biology under construction: in vitro reconstitution of cellular function, Nat Rev Mol Cell Biol, 10 (2009) 644-650.
[23] A. Weinberger, F.C. Tsai, G.H. Koenderink, T.F. Schmidt, R. Itri, W. Meier, T. Schmatko, A. Schroder, C. Marques, Gel-Assisted Formation of Giant Unilamellar Vesicles, Biophys. J., 105 (2013) 154-164.

[24] K.S. Horger, D.J. Estes, R. Capone, M. Mayer, Films of Agarose Enable Rapid Formation of Giant Liposomes in Solutions of Physiologic Ionic Strength, J. Am. Chem. Soc., 131 (2009) 1810-1819.
[25] Rafael B. Lira, R. Dimova, Karin A. Riske, Giant Unilamellar Vesicles Formed by Hybrid Films of Agarose and Lipids Display Altered Mechanical Properties, Biophys. J., 107 (2014) 1609-1619.
[26] T.P.T. Dao, M. Fauquignon, F. Fernandes, E. Ibarboure, A. Vax, M. Prieto, J.F. Le Meins, Membrane properties of giant polymer and lipid vesicles obtained by electroformation and pva gel-assisted hydration methods, Colloids Surf. Physicochem. Eng. Aspects, 533 (2017) 347-353.
[27] D. van Swaay, A. deMello, Microfluidic methods for forming liposomes, Lab Chip, 13 (2013) 752-767.
[28] H. Stein, S. Spindler, N. Bonakdar, C. Wang, V. Sandoghdar, Production of Isolated Giant Unilamellar Vesicles under High Salt Concentrations, Front Physiol, 8 (2017) 63.
[29] T. Portet, R. Dimova, A New Method for Measuring Edge Tensions and Stability of Lipid Bilayers: Effect of Membrane Composition, Biophys. J., 99 (2010) 3264-3273.
[30] C.A. Helm, H. Möhwald, K. Kjaer, J. Als-Nielsen, Phospholipid monolayers between fluid and solid states, Biophys J, 52 (1987) 381-390.
[31] J. Hwang, L.K. Tamm, C. Böhm, T.S. Ramalingam, E. Betzig, M. Edidin, Nanoscale Complexity of Phospholipid Monolayers Investigated by Near-Field Scanning Optical Microscopy, Science, 270 (1995) 610-614.
[32] T. Franke, C. Leirer, A. Wixforth, M.F. Schneider, Phase Transition Induced Adhesion of Giant Unilamellar Vesicles, ChemPhysChem, 10 (2009) 2858-2861.
[33] J. Steinkuhler, J. Agudo-Canalejo, R. Lipowsky, R. Dimova, Variable Adhesion Strength for Giant Unilamellar Vesicles Controlled by External Electrostatic Potentials, Biophys. J., 108 (2015) 402a-402a.
[34] V.D. Gordon, M. Deserno, C.M.J. Andrew, S.U. Egelhaaf, W.C.K. Poon, Adhesion promotes phase separation in mixed-lipid membranes, Europhys. Lett., 84 (2008) 48003.
[35] O. Shindell, N. Mica, M. Ritzer, V.D. Gordon, Specific adhesion of membranes simultaneously supports dual heterogeneities in lipids and proteins, Phys. Chem. Chem. Phys., 17 (2015) 15598-15607.
[36] K. Berndl, J. Kas, R. Lipowsky, E. Sackmann, U. Seifert, Shape transformations of giant vesicles - extreme sensitivity to bilayer asymmetry, Europhys. Lett., 13 (1990) 659-664.
[37] Mark A. Kreutzberger, E. Tejada, Y. Wang, Paulo F. Almeida, GUVs Melt Like LUVs: The Large Heat Capacity of MLVs Is Not Due to Large Size or Small Curvature, Biophysical Journal, 108 (2015) 2619-2622.
[38] N. Bezlyepkina, R.S. Gracià, P. Shchelokovskyy, R. Lipowsky, R. Dimova, Phase Diagram and Tie-Line Determination for the Ternary Mixture DOPC/eSM/Cholesterol, Biophys. J., 104 (2013) 1456-1464.
[39] S.L. Veatch, S.L. Keller, Miscibility phase diagrams of giant vesicles containing sphingomyelin, Phys. Rev. Lett., 94 (2005) 148101.
[40] T.N. Estep, D.B. Mountcastle, Y. Barenholz, R.L. Biltonen, T.E. Thompson, Thermal-behavior of synthetic sphingomyelin-cholesterol dispersions, Biochemistry, 18 (1979) 2112-2117.
[41] P.R. Maulik, G.G. Shipley, N-palmitoyl sphingomyelin bilayers: Structure and interactions with cholesterol and dipalmitoylphosphatidylcholine, Biochemistry, 35 (1996) 8025-8034.
[42] R.F.M. de Almeida, A. Fedorov, M. Prieto, Sphingomyelin/phosphatidylcholine/cholesterol phase diagram: Boundaries and composition of lipid rafts, Biophys. J., 85 (2003) 2406-2416.
[43] A.B. Paul, D.G. Vernita, Z. Zhijun, U.E. Stefan, C.K.P. Wilson, Solid-like domains in fluid membranes, Journal of Physics: Condensed Matter, 17 (2005) S3341.
[44] D. Chen, M.M. Santore, Large effect of membrane tension on the fluid–solid phase transitions of two-component phosphatidylcholine vesicles, Proceedings of the National Academy of Sciences, 111 (2014) 179-184.
[45] Li, J.-X. Cheng, Coexisting Stripe- and Patch-Shaped Domains in Giant Unilamellar Vesicles, Biochemistry-Us, 45 (2006) 11819-11826.

[46] M.A. Johnson, S. Seifert, H.I. Petrache, A.C. Kimble-Hill, Phase Coexistence in Single-Lipid Membranes Induced by Buffering Agents, Langmuir, 30 (2014) 9880-9885.
[47] C.S. Pereira, P.H. Hünenberger, Interaction of the Sugars Trehalose, Maltose and Glucose with a Phospholipid Bilayer: A Comparative Molecular Dynamics Study, The Journal of Physical Chemistry B, 110 (2006) 15572-15581.
[48] R. Dimova, Recent developments in the field of bending rigidity measurements on membranes, Adv. Colloid Interface Sci., 208 (2014) 225-234.
[49] J.F. Nagle, Introductory Lecture: Basic quantities in model biomembranes, Faraday Discuss., 161 (2013) 11-29.
[50] J.F. Nagle, M.S. Jablin, S. Tristram-Nagle, K. Akabori, What are the true values of the bending modulus of simple lipid bilayers?, Chemistry and physics of lipids, 185 (2015) 3-10.
[51] C.G. Sinn, M. Antonietti, R. Dimova, Binding of calcium to phosphatidylcholine-phosphatidylserine membranes, Colloids and Surfaces a-Physicochemical and Engineering Aspects, 282 (2006) 410-419.
[52] K. Oglecka, J. Sanborn, A.N. Parikh, R.S. Kraut, Osmotic Gradients Induce Bio-reminiscent Morphological Transformations in Giant Unilamellar Vesicles, Frontiers in Physiology, 3 (2012).
[53] K. Oglęcka, P. Rangamani, B. Liedberg, R.S. Kraut, A.N. Parikh, Oscillatory phase separation in giant lipid vesicles induced by transmembrane osmotic differentials, eLife, 3 (2014) e03695.
[54] H.D. Andersen, C.H. Wang, L. Arleth, G.H. Peters, P. Westh, Reconciliation of opposing views on membrane-sugar interactions, Proc. Natl. Acad. Sci. U. S. A., 108 (2011) 1874-1878.
[55] D. Weinbuch, J.K. Cheung, J. Ketelaars, V. Filipe, A. Hawe, J. den Engelsman, W. Jiskoot, Nanoparticulate Impurities in Pharmaceutical-Grade Sugars and their Interference with Light Scattering-Based Analysis of Protein Formulations, Pharmaceutical Research, 32 (2015) 2419-2427.
[56] P. Carravilla, J.L. Nieva, F.M. Goni, J. Requejo-Isidro, N. Huarte, Two-Photon Laurdan Studies of the Ternary Lipid Mixture DOPC:SM:Cholesterol Reveal a Single Liquid Phase at Sphingomyelin:Cholesterol Ratios Lower Than 1, Langmuir, 31 (2015) 2808-2817.
[57] B. Kubsch, T. Robinson, R. Lipowsky, R. Dimova, Solution Asymmetry and Salt Expand Fluid-Fluid Coexistence Regions of Charged Membranes, Biophys. J., 110 (2016) 2581-2584.

**Figure Captions**

**Figure 1. Chamber used for temperature control and vesicle observation.** More details on assembling the chamber as well as photos, can be found in Ref. [20].

**Figure 2. Domains in gel-phase DPPC giant vesicles after a heating/cooling treatment.** Two GUVs (labeled with Rh-DPPE) prepared at 60 °C, cooled at a rate of 0.15 K/min to room temperature, diluted in glucose, and then reheated above the phase transition temperature (to 60 °C) followed by cooling to room temperature and imaging: (A) equatorial cross section; (B) confocal image of the upper parts of the vesicles and (C) phase contrast are shown: the left vesicle has leaked as judging from the lost contrast, while the membrane of the right vesicle is not compromised. Scale bar: 20 μm. (D) Line profile across the domain indicated with a yellow line in (B) shows that the domain diameter is approximately 3 μm.

**Figure 3. Formation of dye-depleted domains and stripe-like structures in DPPC GUVs during cooling.** The GUVs were prepared in 200 mM sucrose, slowly cooled down to room temperature, diluted in glucose, reheated above $T_M$ and observed with confocal microscopy while cooling (for cooling rate see C). (A, B) The images show confocal snapshots at different times as illustrated in panel (C). They were recorded at the bottom of the chamber where the vesicles are almost flat and convenient for imaging (similar behavior was observed on the upper pole of the GUV, not adhering to the substrate). (A) Fluid black domains nucleation and growth several degrees above the main phase transition temperature (see panel (C)). Time zero refers to the last frame before domains were detected. The arrowheads point to domains with heart-like shapes and arrows to kinks at the domain boundaries. (B) Appearance and growth of stripe-like structure, enriched in dye, preferentially accumulated at the rim of the black domains. Time zero refers to the last frame before stripe-like structures were observed. (C) Temperature decay in the experimental chamber with time (n=1). The light grey area indicates the times, when domains are observed to nucleate as shown in (A). The dark grey area shows the period when stripe-like structures form as shown in (B). The main phase transition temperature of DPPC GUVs $T_M$(GUVs) = 41.7°C following [37] is indicated with the dashed line and the hatched region around it shows the corresponding half-width of the transition. Scale bars: 20 µm.

**Figure 4. Reversible domain formation and domain fluidity in DPPC GUVs.** (A) DPPC GUVs (same vesicles on all snapshots) recorded at 52°C, cooled to 49 °C, and reheated to 52 °C. Small domains in the lower part of the vesicles are observed at 49 °C (more clearly seen in the larger vesicle with zoomed insert, see Fig. S6 for additional micrographs of the upper part of this vesicle). (B) Time series showing domain coalescence at 51°C. Initially, black domains that appear circular are free to diffuse over the GUV surface, over time domains start to fuse and show a slow relaxation towards more rounded shape. Images obtained by epifluorescence microscopy. Scale bars 10 µm.

**Figure 5. Domains in quasi single-component vesicles made of different phosphatidylcholines.** The vesicles were prepared from (A) DSPC, (B) DPPC and (C) SOPC, as explained in the text, and cycled across the main phase transition temperature of the respective lipid. All images were taken at the bottom of the experimental chamber after equilibration at room temperature. In these conditions, the vesicles made of DSPC and DPPC are in the gel phase and exhibit domains with facets (A, B), whereas SOPC vesicles are in the fluid phase and the domains have smooth boundaries (C). Scale bars: 10 µm.

**Figure 6. Formation of micron-sized domains is suppressed by adding lipids in mM concentration.** Surface patterns on GUVs prepared in 0.2 M sucrose, cooled down and diluted 1:1 with (A) 0.2 M glucose, (B) 0.2 M glucose containing 1 mM SOPC SUVs, and (C) 0.2 M glucose containing 10 mM SOPC SUVs (employing DPPC SUVs gave similar results). The vesicles were then heated to 60 °C, cooled down and imaged. Confocal section of the vesicle poles were acquired at room temperature. Scale bars 10 µm.

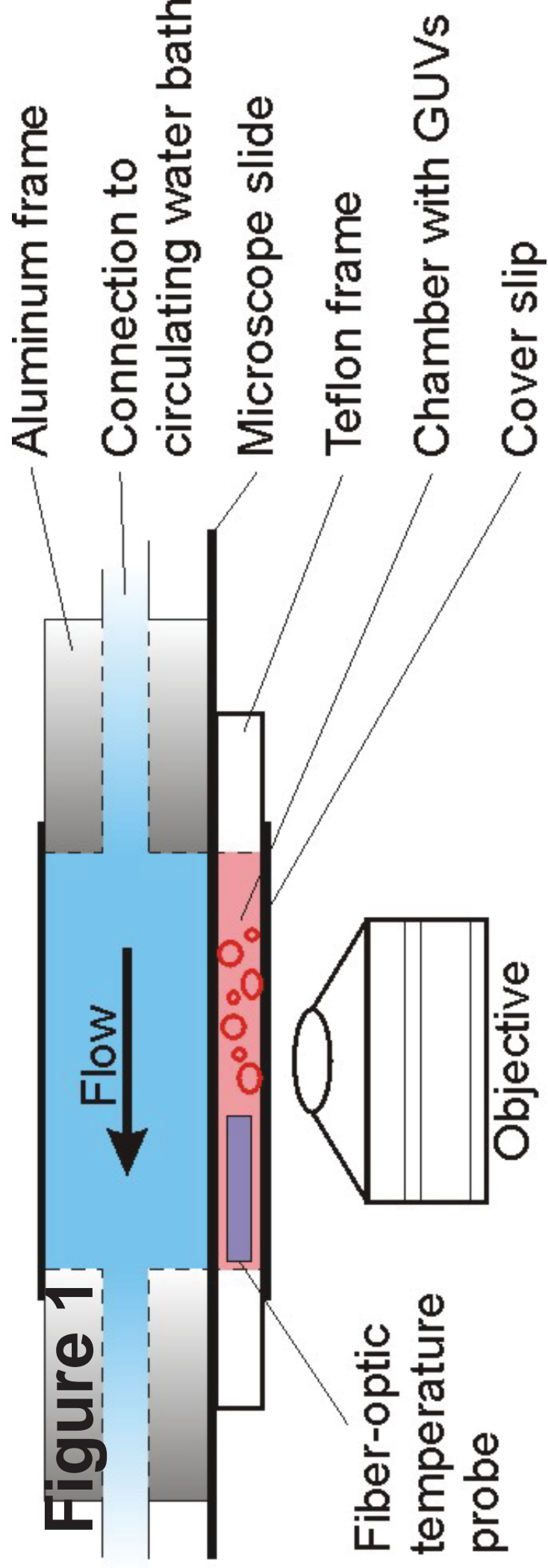

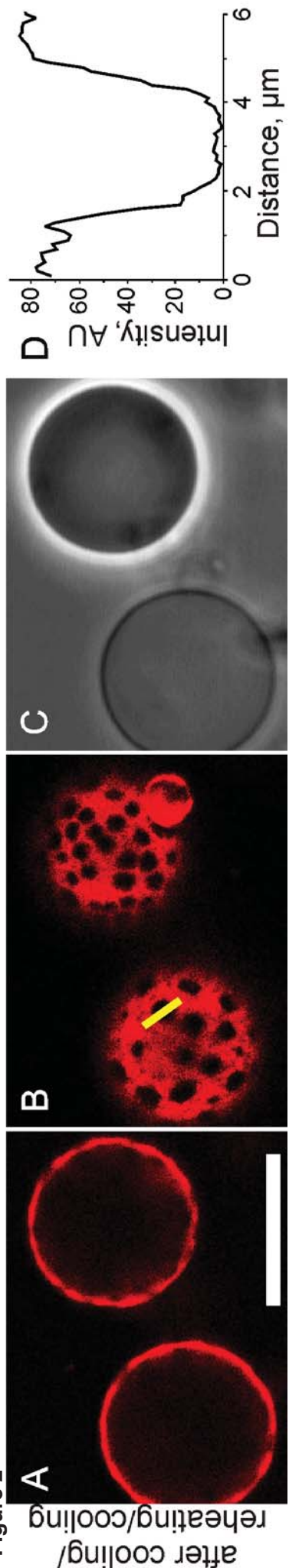

Figure 2

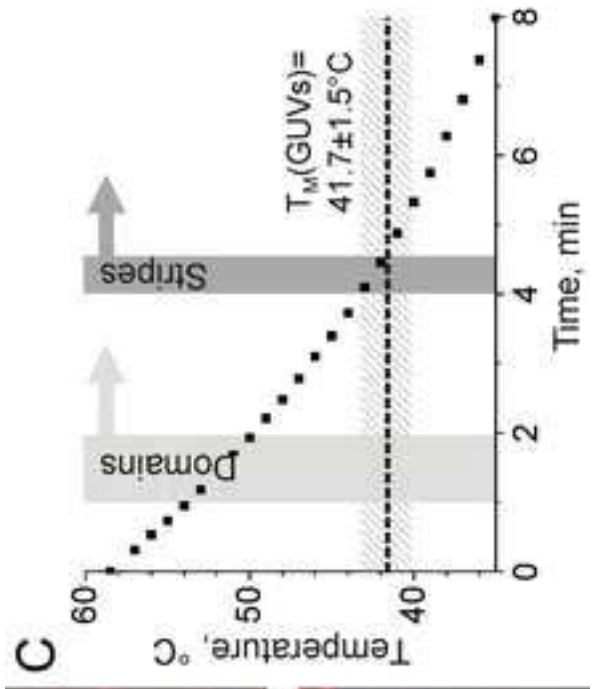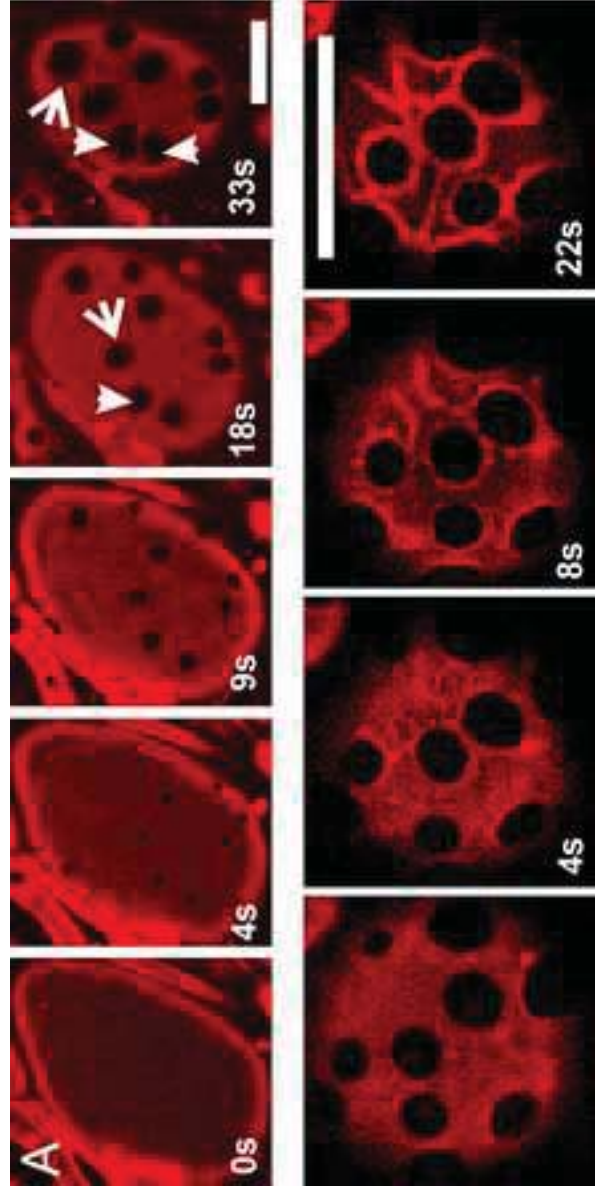

Figure 4

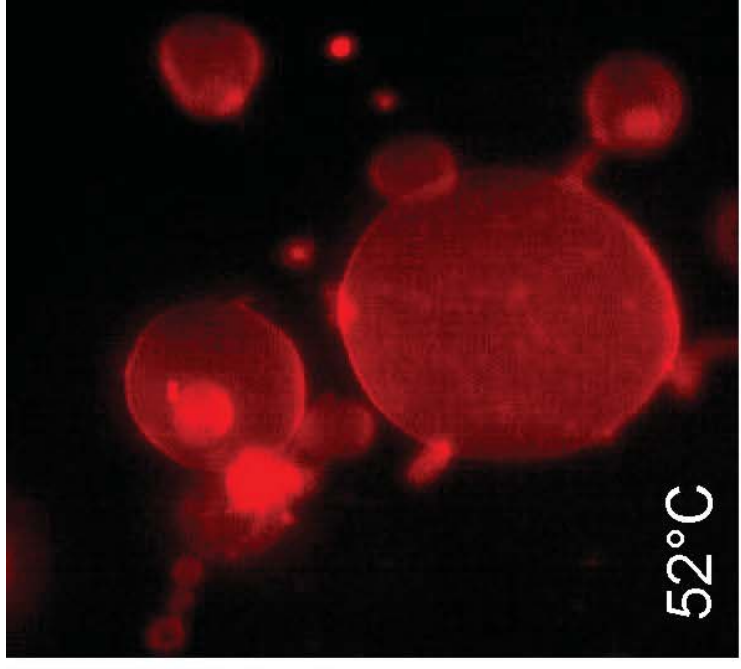
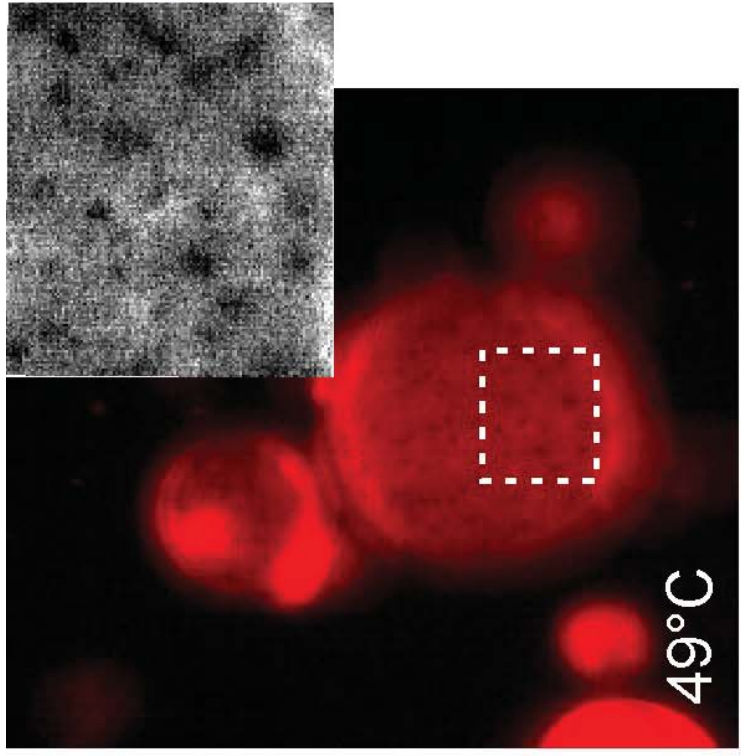
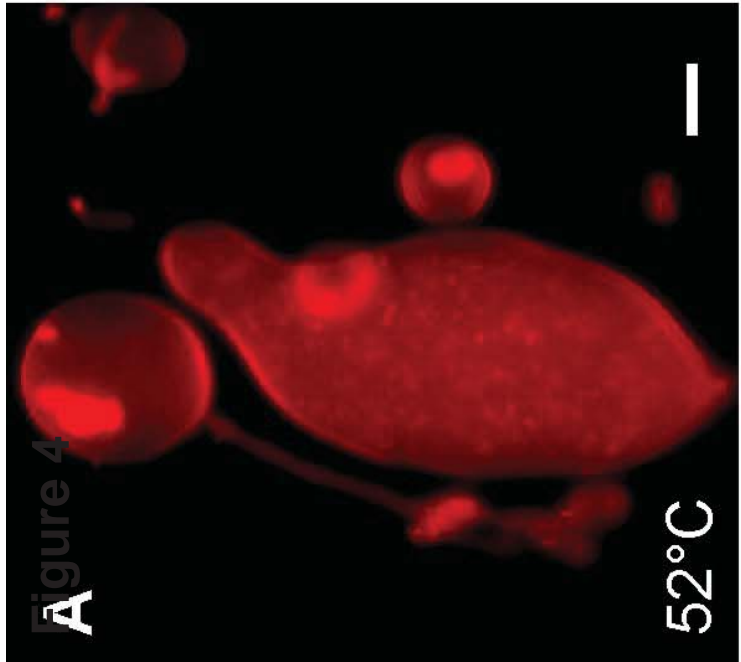

A  52°C  49°C  52°C

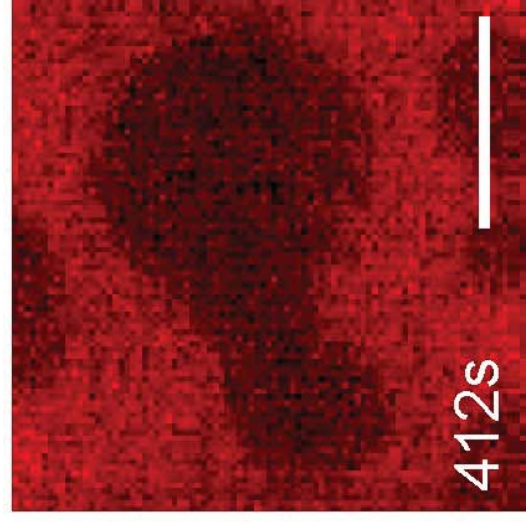
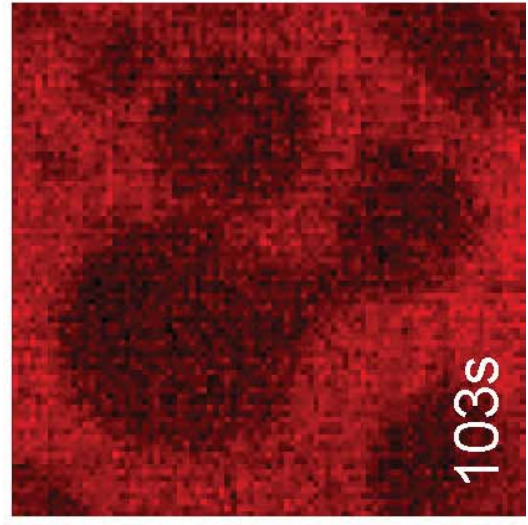
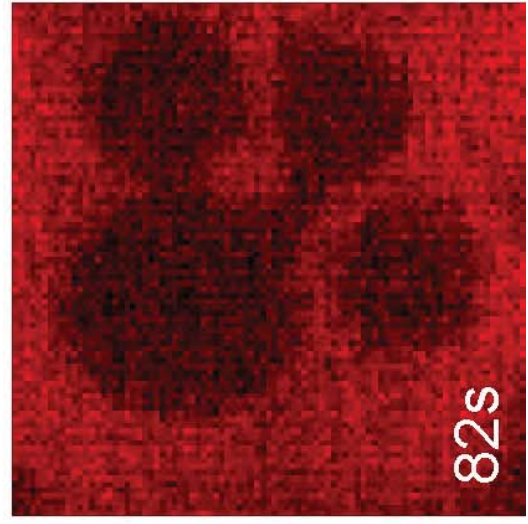
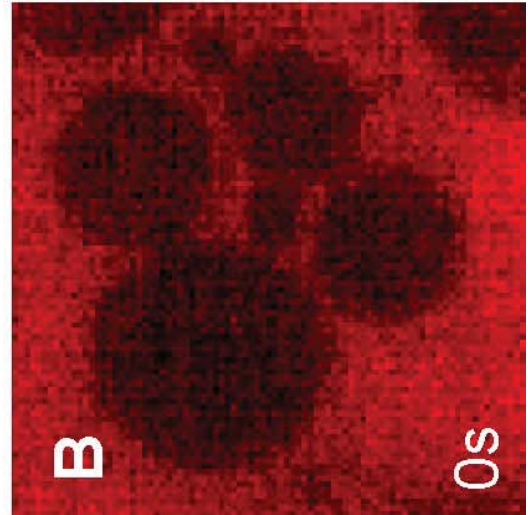

B  0s  82s  103s  412s

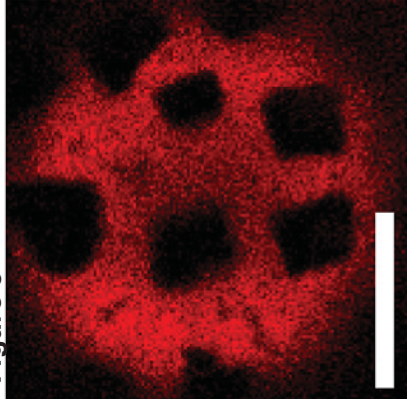
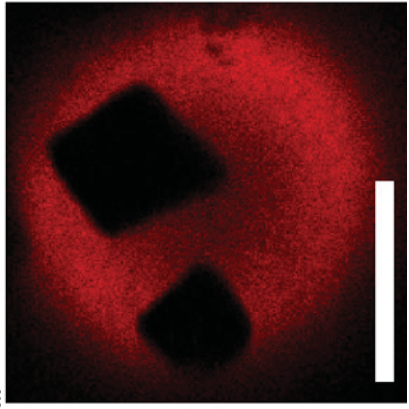
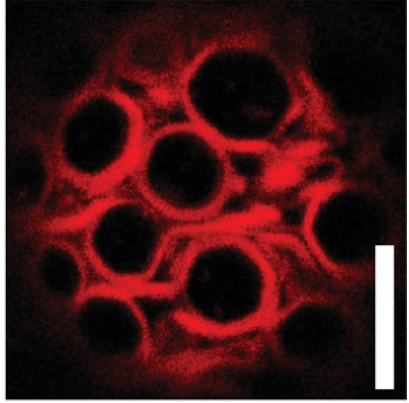
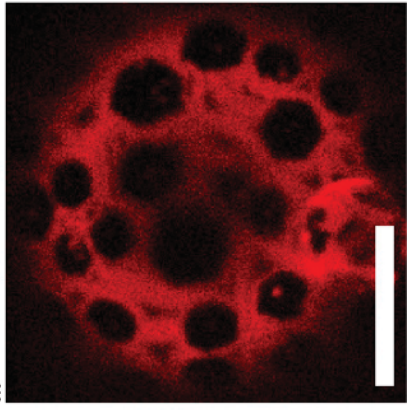
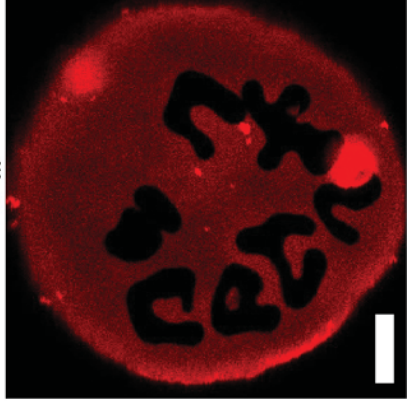

Figure 5

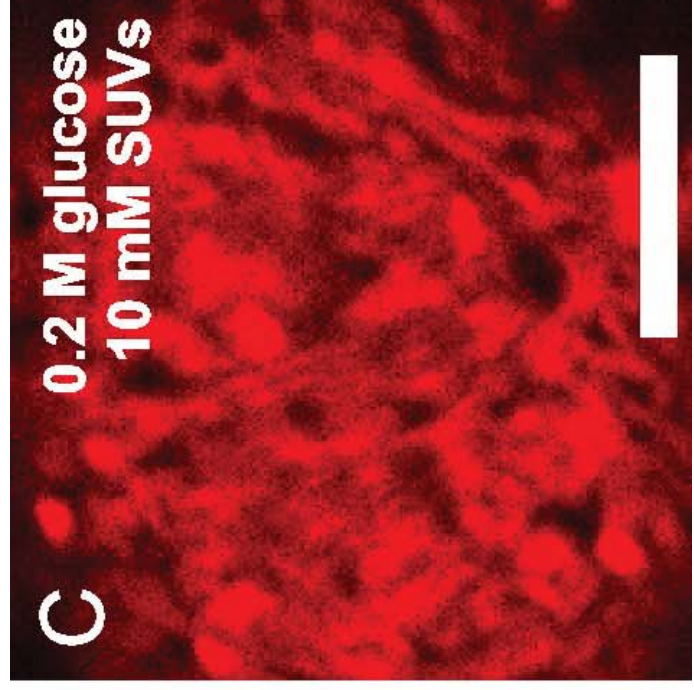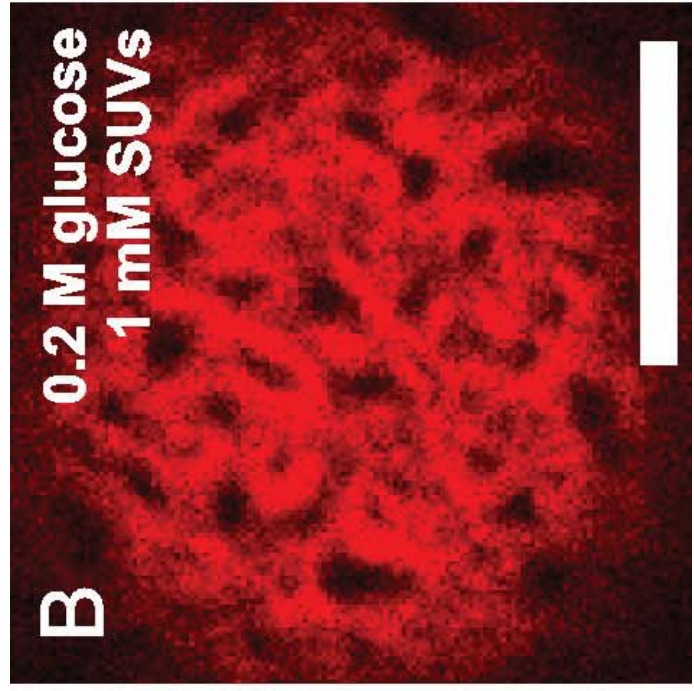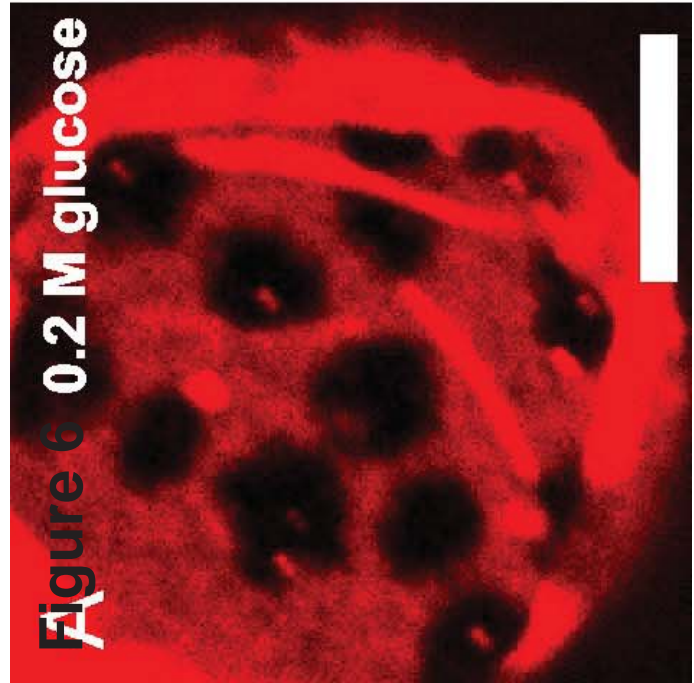

Figure 6  0.2 M glucose | 0.2 M glucose 1 mM SUVs | 0.2 M glucose 10 mM SUVs

# Supporting Information

# Micron-sized domains in quasi single-component giant vesicles

Roland L. Knorr*, Jan Steinkühler and Rumiana Dimova*

Max Planck Institute of Colloids and Interfaces, Department of Theory & Bio-Systems, Science Park Golm, 14424 Potsdam, Germany

*Address correspondence to: knorr@mpikg.mpg.de and dimova@mpikg.mpg.de

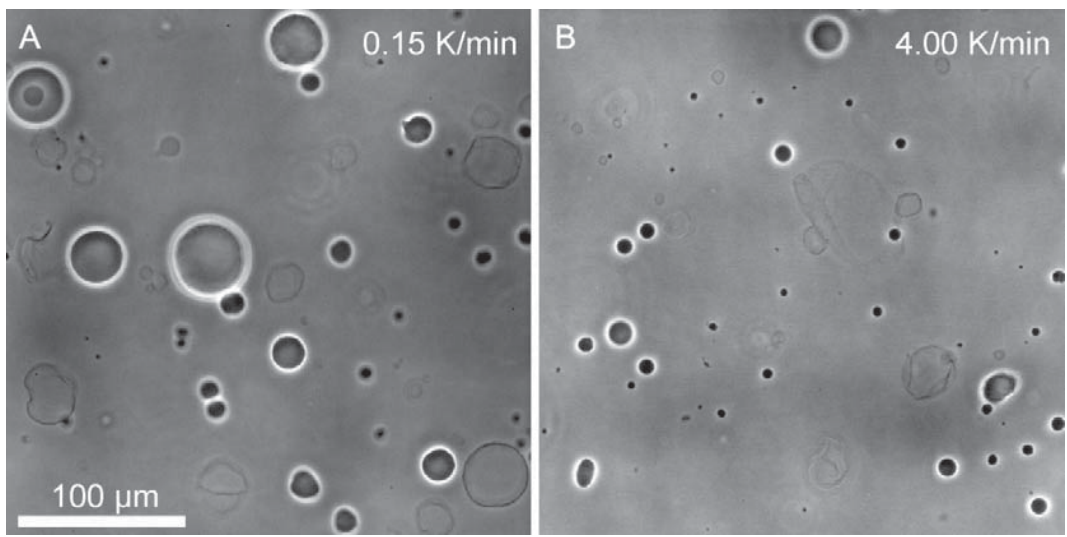

**Figure S1. Slow cooling improves the contrast and stability of GUVs in the gel phase.** The vesicles were grown in 200 mM sucrose at 60 °C, cooled down at different rates to room temperature as indicated, diluted with isotonic glucose solution and observed in phase contrast. (A) Upon slow cooling of the sample (0.15 K/min), a large fraction of the vesicles survive without leaking as demonstrated from the preserved phase contrast. (B) Upon fast cooling of the GUVs (4 K/min), only small vesicles with persevered contrast remain.

## Section S1. Heterogeneities in the membrane surface of DPPC-GUVs; heterogeneities do not result from the preparation method and solutions

After preparation and slow cooling, the vesicles were examined with confocal microscopy. 3D confocal projections of GUVs showed inhomogeneous fluorescence over the vesicle surface, Fig. S2. The observed structures are reminiscent of simulation snapshots of gel phase membranes (see e.g. [1, 2]) although at a very different scale. The fluorophore in the GUVs (in this case Rh-DPPE) appeared inhomogeneously distributed over the vesicle surface. Such inhomogeneous fluorescence intensity can be caused by dye partitioning and local membrane corrugations. Note that this inhomogeneity is difficult to detect with epifluorescence imaging because of the poorer resolution compared to confocal microscopy. The inhomogeneity is observed on GUVs both with preserved and lost phase contrast, and on vesicles with smooth surface (as in Fig. 2) or strongly corrugated ones as in Fig. S3D,E.

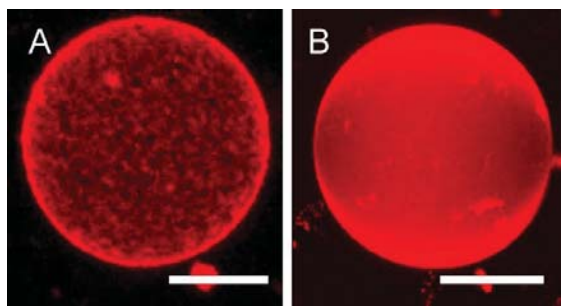

**Figure S2. Confocal 3D projection of vesicles.** (A) DPPC vesicle after preparation in 200 mM sucrose, slow cooling and dilution in isotonic glucose solution (image acquired at room temperature). The membrane surface exhibits grainy pattern of the distribution of the membrane dye (Rh-DPPE). (B) For comparison, a vesicle in the fluid phase with homogeneous surface. Scale bars: 20 µm.

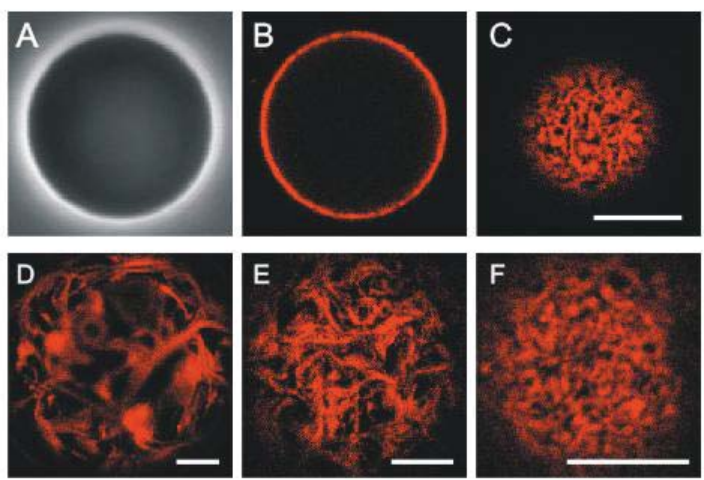

**Figure S3. Inhomogeneity in the membrane of DPPC GUVs after electroformation and slow cooling to room temperature.** (A-C) Images of one GUV (labeled with Rh-DPPE) after electroformation in 200 mM sucrose solution at 60 °C, slow cooling (0.15 K/min) to room temperature and dilution in isotonic glucose solution at 1:10 ratio: (A) phase contrast, (B) equatorial cross section of the same vesicle. (C) Confocal image of the upper part of the vesicle where the grainy pattern can be more clearly observed. (D-F) Additional examples of surface corrugations observed on vesicles with visibly wrinkled surface; the GUV shown in (D) was in a sample not diluted in glucose and is strongly corrugated. Scale bars: 10 µm.

We set to explore whether the observed graininess in fluorescence intensity is related to our particular mode of observation and the vesicle preparation protocol. Possible artifacts that could arise from light-induced domain formation [3, 4] can be excluded, as the vesicles were in the gel phase and the surface inhomogeneity in the membrane persisted even for long illumination times. In addition, illumination with low intensity was used for the recordings. We examined the effect of sucrose (note that compared to the total lipid concentration in the electroformation chamber of 40 µM, sucrose is in strong excess at 200 mM concentration). However, electroformation at lower sucrose concentration (20 mM) did not change the outcome. The effect of glucose could be excluded as well, as samples without dilution in glucose showed similar structures, see Fig. S3D. This outcome was expected, since the membrane is already in the gel phase at room temperature and simple dilution with glucose should not alter the distribution of dyes on our experimental time scale. Indeed, we can entirely exclude the contribution from sugars, because the inhomogeneity in the membrane was observed also in GUVs grown in pure water.

We then questioned the effect of the electroswelling protocol. Our electroformation conditions were relatively mild to expect oxidation effects as those reported in Refs. [5]. This is understandable as the acyl chains of DPPC are saturated. Effects associated with ITO electrodes [3] can also be excluded as formation on platinum wires showed similar results. Finally, vesicles prepared by spontaneous swelling in 200 mM sucrose behaved in the same way, which entirely excludes artifacts associated with electroformation and sugars. The examined conditions are summarized in Table S1.

Deflation and handling of suspensions of giant vesicles in the gel phase can give rise to surface corrugations, faceted shapes or topological defects. The latter can result from rough manipulation of the solutions or can be due to an interplay between the melting and freezing behavior of the lipids and mechanical constrains of the vesicles [1, 6-9]. Thus, we tested whether the patterns will smooth out or get enhanced at room temperature by inflating and deflating the GUVs slightly by applying hyper- or hypotonic conditions (+/- 10 osmol%) instead of isosmotic dilution. Such a treatment, specifically hypotonic solutions, can be expected to fully inflate nearly spherical vesicles and thus, to flatten out corrugations. However, the inhomogeneity of the vesicle surface remained unchanged. We thus conclude that the inhomogeneous structure of the membrane is not an artifact of the preparation method. We did not explore this phenomenon further but focused on understanding the appearance of the micron-sized domains as observed in Figs. 2 and S4.

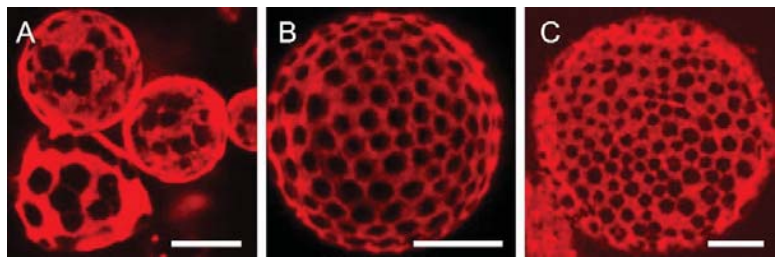

**Figure S4. Examples of vesicles with domains after a reheating cycle.** (A, B) 3D projection of confocal sections. (C) Single confocal section of a flat part of a GUV in contact with the chamber bottom. Images were acquired after cooling to room temperature. Scale bars: 10 µm.

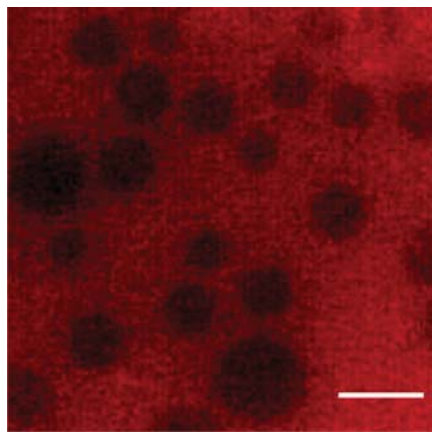

**Figure S5. Domain pattern on DPPC GUV that develops over time (approximately 5 minutes after crossing the phase transition temperature) at constant temperature of 49°C.** Scale bar indicates 5 µm.

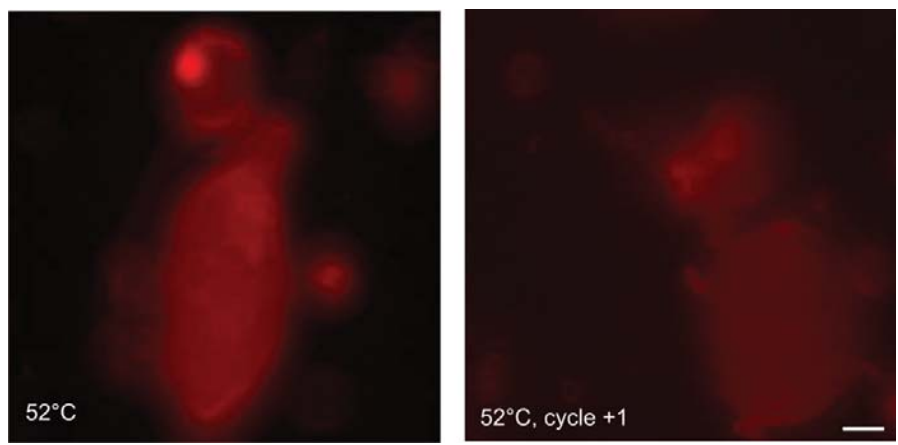

**Figure S6. Epifluorescence microscopy images of the upper hemisphere of the same DPPC GUV shown in Figure 4 before (left) and after a cooling cycle (right).** The focus was slightly adjusted to display the upper part of the vesicle. Both images indicate no (black)-domain formation. Scale bar indicates 10 μm.

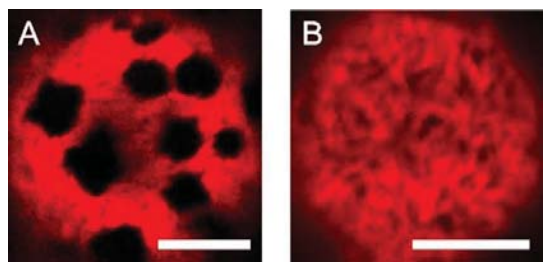

**Figure S7. Dye-depleted micron-sized domains do not form upon co-incubation with an excess of lipids.** Experiment as in Fig. 6 in the main text but at reduced concentration of sucrose/glucose. The vesicles were grown in 15 mM sucrose, cooled down and diluted 1:10 with (A) 15 mM glucose or (B) 15 mM glucose containing 1 mM SOPC SUVs. The vesicles were then reheated above $T_M$. Cooled to room temperature, the vesicle poles were observed by confocal microscopy. Scale bars: 10 μm.

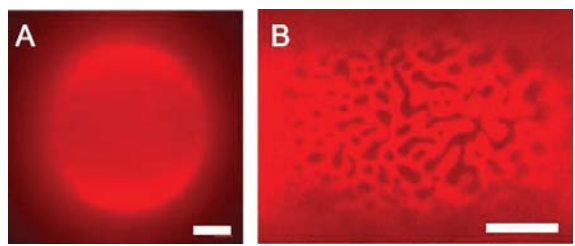

**Figure S8. Formation of domains in 7/3 eSM:Chol membranes in the presence of glucose.** Epifluorescence micrographs of (A) a GUV grown in pure water, and (B) GUV grown from the same mixture in 50 mM sucrose and diluted in isotonic glucose. The focal plane is set to the upper pole of the GUVs. Images were recorded at room temperature. Scale bars: 10 μm.

**Table S1. Summary of the experimental conditions used in this study.**

| Main GUV lipid | GUV swelling solution | Solution for GUV dilution (at 22°C) | Microscopic domains after cycling across $T_M$ | Other conditions tested |
|---|---|---|---|---|
| L-DPPC | Sucrose 0.2M | Glucose 0.2M | yes | electroformation on ITOs or Pt-wires, spontaneous swelling, various dyes: NBD-DPPE, TR-DHPE, perylen |
| | Sucrose 0.2M | Glucose 0.22M | yes | |
| | Sucrose 0.2M | Glucose 0.18M | yes | |
| | Sucrose 0.02M | Glucose 0.02M | yes | |
| | Sucrose 0.2M | no dilution | no | |
| | Bidest. water | no dilution | no | |
| | Sucrose 0.2M | Glucose 0.2M with 10 mM lipid (SUVs) | no | SUVs made of DPPC and SOPC |
| | Sucrose 0.02M | Glucose 0.02M with 1 mM lipid (SUVs) | no | |
| D/L-DPPC | Sucrose 0.2M | Glucose 0.2M | yes | |
| D-DPPC | Sucrose 0.2M | Glucose 0.2M | yes | |
| L-DSPC | Sucrose 0.2M | Glucose 0.2M | yes | |
| L-SOPC | Sucrose 0.2M | Glucose 0.2M | yes | |
| eSM:Chol = 7/3 | Bidest. water | no dilution | no | |
| eSM:Chol = 7/3 | Sucrose 0.05 M | Glucose 0.05 M | Domains formed without cycle | |

**References**

[1] J.H. Ipsen, K. Jørgensen, O.G. Mouritsen, Density fluctuations in saturated phospholipid bilayers increase as the acyl-chain length decreases, Biophysical Journal, 58 (1990) 1099-1107.

[2] James A. Svetlovics, Sterling A. Wheaten, Paulo F. Almeida, Phase Separation and Fluctuations in Mixtures of a Saturated and an Unsaturated Phospholipid, Biophys. J., 102 (2012) 2526-2535.

[3] A.G. Ayuyan, F.S. Cohen, Lipid peroxides promote large rafts: Effects of excitation of probes in fluorescence microscopy and electrochemical reactions during vesicle formation, Biophys. J., 91 (2006) 2172-2183.

[4] J. Zhao, J. Wu, H.L. Shao, F. Kong, N. Jain, G. Hunt, G. Feigenson, Phase studies of model biomembranes: Macroscopic coexistence of L alpha plus L beta, with light-induced coexistence of L alpha plus Lo Phases, Biochim. Biophys. Acta-Biomembr., 1768 (2007) 2777-2786.

[5] M. Breton, M. Amirkavei, L.M. Mir, Optimization of the Electroformation of Giant Unilamellar Vesicles (GUVs) with Unsaturated Phospholipids, J Membrane Biol, 248 (2015) 827-835.

[6] F. Quemeneur, C. Quilliet, M. Faivre, A. Viallat, B. Pépin-Donat, Gel Phase Vesicles Buckle into Specific Shapes, Phys Rev Lett, 108 (2012) 108303.

[7] T. Franke, C. Leirer, A. Wixforth, M.F. Schneider, Phase Transition Induced Adhesion of Giant Unilamellar Vesicles, ChemPhysChem, 10 (2009) 2858-2861.

[8] H.V. Ly, M.L. Longo, Probing the Interdigitated Phase of a DPPC Lipid Bilayer by Micropipette Aspiration, Macromolecular Symposia, 219 (2005) 97-122.

[9] L.S. Hirst, A. Ossowski, M. Fraser, J. Geng, J.V. Selinger, R.L.B. Selinger, Morphology transition in lipid vesicles due to in-plane order and topological defects, Proceedings of the National Academy of Sciences, 110 (2013) 3242-3247.